# Validation of a propulsion model in front crawl swimming

Thomas Brunel[1], Charlie Prétot[1], Christophe Clanet[2], Baptiste Bolon[1], Frédérique Larrarte[1,3], Caroline Cohen[2] & Rémi Carmigniani[1]
[1]LHSV, École des Ponts, EDF R et D, UPE, Chatou, France, [2]LadHyX, UMR 7646 du CNRS, Ecole Polytechnique, Palaiseau, France, [3]Univ. Gustave Eiffel, Marne la Vallée, France.

## INTRODUCTION

In swimming competition, among all the swimming style, front crawl is the one which allows to reach the highest speed and appear to be the most efficient (Barbosa, et al., 2010).

Thus, it is the stroke use in all the freestyle events. It covers a range of distance from 50 m to 1500 m in swimming pool and from 5 km to 25 km for open water swimming.

Front crawl swimming is characterized by an alternate action of the arms. A cycle is composed of three main phases for each arm: a gliding phase just after the hand entry into the water with one arm extended forward; a propulsion phase (pull and push) and a recovery phase. The frequency of this cycle is commonly called stroke rate. The study of the relationship between velocity and stroke rate leads to focus on arm organization during a cycle (Chollet, Chalies, & Chatard, 2000). Chollet *et al.* (2000) highlighted that there are different cycle coordinations depending on the swimming speed. At low velocities, typical of races longer than 200 meters, swimmers mark a gliding pause with one arm extended forward during their cycle before a propulsion phase. This coordination is called *catch-up* mode. As the pace increases, the gliding pauses become shorter, and the propulsion phases become dominant. Some elite swimmers are even able to superpose the propulsion phases of the two arms using a fast recovery (Seifert, Chollet, & Rouard, 2007). This coordination is called *superposition* mode.

In their work, Carmigniani *et al.* (Carmigniani, Seifert, Chollet, & Clanet, 2020) present a propulsion model to explain the general evolution of the coordination based on the minimization of energy during a cycle. This model outlines two regimes. The first one at low speed where the swimmers vary their speed with the force they use per stroke and maintain constant coordination. The second one appears when a critical velocity is reach which means that the swimmers are at maximum force. In order to further increase the swim velocity, they start to reduce the gliding and recovery phases.

In this study, we propose a first validation of this model through a progressive speed test, using instrumented paddles to measure the force generated by the arms of the swimmers and a video recording system to measure the velocity.

## METHODS

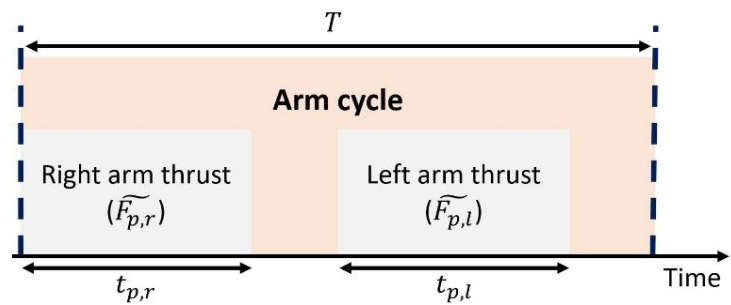

*Figure 1 : Temporal scheme of an arm cycle. It starts with the beginning of the thrust of the right arm and finish with the beginning of the next thrust of this same arm. T is the period of the cycle and $t_{p,r}$ ($t_{p,l}$) is the propulsion time during which the right (left) arm generates a thrust $\widetilde{F_{p,r}}$ ($\widetilde{F_{p,l}}$, respectively).*

One can describe an arm cycle as following. It starts with the beginning of the propulsion of one arm and finish with the beginning of the next propulsion of this arm (Figure 1). $T$ is the period of the cycle and $t_{p,r}$ ($t_{p,l}$) is the propulsion time during which the right (left) arm generates a thrust $\widetilde{F_{p,r}}$ ($\widetilde{F_{p,l}}$, respectively).

By writing Newton's Second Law on the swimmer system in the direction of the race and averaging on a stroke cycle, Carmigniani et al.(2020) shows that the equilibrium of forces gives:

$$\frac{1}{T}\int_0^T F_p(t)dt = k_b v^2,$$

(1)

With $F_p(t)$ the propulsion force of the swimmer over time, $k_b$ the active drag coefficient (supposed constant) and $v$ the average velocity. The expression $\overline{F} = \frac{1}{T}\int_0^T F_p(t)dt$ is the propulsive term of the equation and represent the mean force generated by the swimmer. One can suppose that this propulsive term is mainly due to arms actions or at least highly corelated to it. If $t_{start,r}$ ($t_{start,l}$) is the beginning of the propulsion and $t_{end,r}$ ($t_{end,l}$) the end of the propulsion of the right (left) arm, the propulsion time of the arm $t_{p,r}$ ($t_{p,l}$, respectively) could be write as $t_{p,r} = t_{end,r} - t_{start,r}$ ($t_{p,l} = t_{end,l} - t_{start,l}$). The propulsive term becomes then:

$$\overline{F} = \frac{1}{T}\left\{\int_{t_{start,r}}^{t_{end,r}} F_p(t)dt + \int_{t_{start,l}}^{t_{end,l}} F_p(t)dt\right\}.$$

(2)

Hence it comes:

$$\overline{F} = \frac{t_{p,r}}{T}\widetilde{F_{p,r}} + \frac{t_{p,l}}{T}\widetilde{F_{p,l}},$$

(3)

where $\widetilde{F_{p,r}} = \frac{1}{t_{p,r}}\int_{t_{start,r}}^{t_{end,r}} F_p(t)dt$ ($\widetilde{F_{p,l}} = \frac{1}{t_{p,l}}\int_{t_{start,l}}^{t_{end,l}} F_p(t)dt$) is the mean propulsion force of the right (left, respectively) arm. For the sake of simplicity, one can do the hypothesis of symmetry between right and left arm. Then, $t_{p,r} = t_{p,l} = t_p$ and $\widetilde{F_{p,r}} = \widetilde{F_{p,l}} = \widetilde{F_p}$ which allow to express the propulsive term as:

$$\overline{F} = \frac{2t_p}{T}\widetilde{F_p}.$$

(4)

Here appears the relative propulsion time $i_{prop} = \frac{2t_p}{T}$. Therefore, equation (1) could be rewrite as:

$$i_{prop}\widetilde{F_p} = k_b v^2.$$

(5)

Two regimes are found from this equation, the gliding regime and the maximum force regime. The gliding regime is characterized by a constant relative propulsion time $i_{prop}$ and an increase of the swimming velocity with the increase of the thrust $\widetilde{F_p}$. The transition between the gliding regime and the maximum force regime happens at a critical speed where the swimmer reaches his maximal thrust. Here, the thrust becomes constant and maximum ($\widetilde{F_p} = F^*$) and the swimmers further increases their velocity $v$ by increasing the relative propulsion time $i_{prop}$.

Variation of $i_{prop}$ and $\widetilde{F_p}$ were studied during a 5x25 m test with progressive speed by 25 m. No feedbacks were given during the test and the swimmers had three minutes rest between each length in order to avoid any effect of fatigue.

The test is performed with instrumented swimming paddles equipped with synchronised force sensors and inertial motion units. It allows to measure the force normal to the paddle's plane, the angular velocity and the acceleration. The sessions were recorded using 5 underwater and 5 aerial fixed cameras (Figure 2).

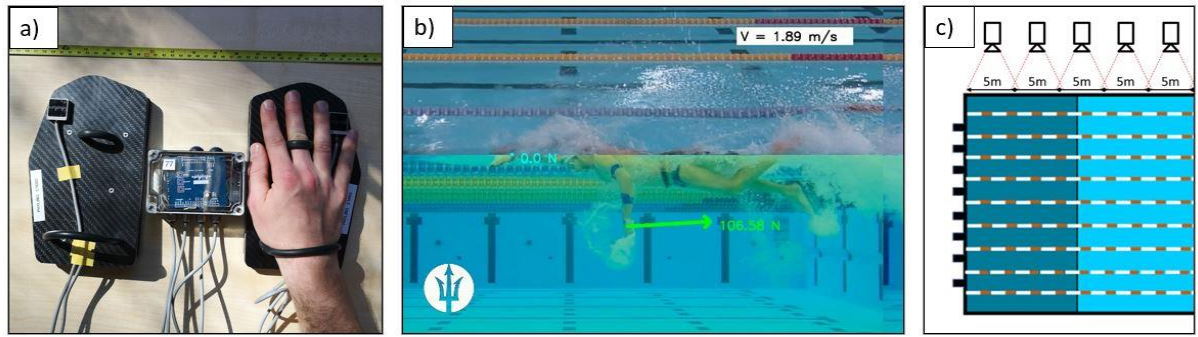

*Figure 2 : a) Instrumented swimming paddles. b) Fusion of the data from the video recording and the instrumented swimming paddles. c) Scheme of the swimming pool with the 5 aerials cameras covering the first 25m. 5 other cameras are in place for underwater footage.*

The test session takes place as follow: swimmers arrive at the swimming pool and will have 20 min to warm up. Then, go out of the water and will be equipped with the instrumented paddles (Figure 3, a). Time will be given to them to get used of swimming with the device. After that, the 5x25 m test with an increasing speed by length can start. The video recording and the instrumented paddles are launched. Before each pool length, the following steps are performed by the swimmer:
- a "clap" is done with the paddles to synchronise the video and the paddles,
- the hands are kept in the air for 5 seconds to initialise the force sensors in the paddles,
- the hands are puts in paddles print of known orientation for 10 seconds to initialise the orientation of the paddles (Figure 3, b),
- once the 25 m have been reached, they come back to the start and wait for the end of the 3 minutes rest before the next length.

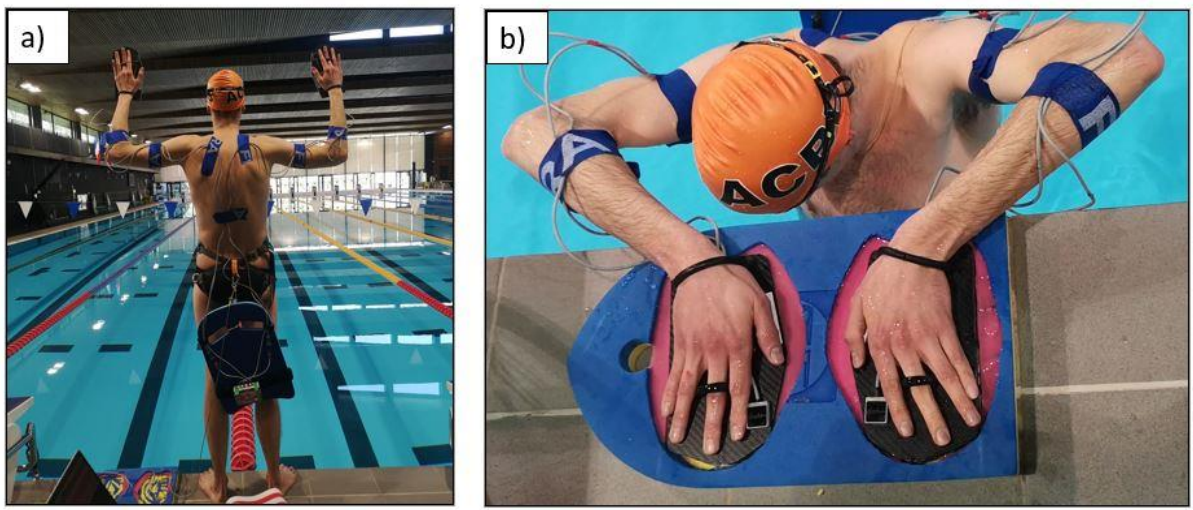

*Figure 3 : a) Swimmer equipped with the instrumented paddles. b) Initialisation of the orientation of the paddles.*

All the data from the experiments are processed with a Python code. The cameras are calibrated in 3D thanks to the Open CV library (OpenCV, n.d.) with less than 1 cm of error on the accuracy of reprojection. A homemade neural network with a U-net architecture have been trained to follow remarkable points of the human body (head, shoulders, elbows, wrists, hips, knees, ankles) while swimming. The position of the markers is used to evaluate the position of the centre of mass from de Leva tables (Leva, 1996). The accuracy of the centre of mass position is about 3 cm. The obtained data are validated by hands by an operator. It finally allows to compute their velocity $v$.

Because the thrust $\tilde{F}_p$ is aligned with the swimming direction, it is necessary to project the measured force in this direction. This implies to know the orientation of the paddle over time. A Madgwick filter algorithm (Madgwick), (Kadi, et al., 2022), is used for the purpose. Once the thrust over time for each arm is obtained, one could determine the propulsion time and the period of a cycle to compute the relative propulsion time $i_{prop}$.

Two swimmers were tested. N1 is a male swimmer competing at national level and N2 is a male swimmer competing at international level.

**RESULTS**

The evolution of the velocity and forces over time is extracted from the experiment for each pool length (Figure 4, a). Based on the projected forces and the video, the beginning ($t_{start,r}$, $t_{start,l}$) and the end ($t_{end,r}$, $t_{end,l}$) of the impulsions of each arm in each cycle are identified (Figure 4, b). Then, the mean force $\bar{F}$, the propulsion time $t_p$, the period $T$ and the mean velocity $v$ are obtained by being averaged over three cycles. The range of obtained velocities is from 1.29 m/s to 1.91 m/s for N1 and from 1.04 m/s to 1.81 m/s for N2.

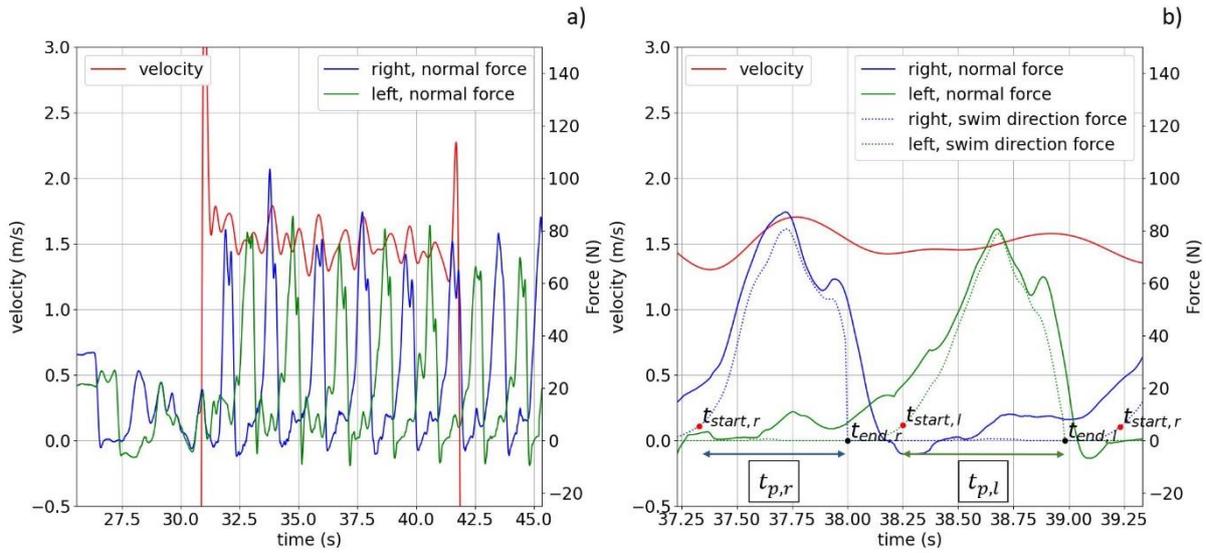

*Figure 4 : Evolution of velocity and forces over time for N2, length 3. a) Full 25m, evolution of the velocity and normal forces. b) Zoom on a cycle, comparison between normal forces and projected forces along the swim direction.*

Knowing the mean force and the mean velocity for every length, one can look after the equation ( 1) and rewrite it as follow:

$$v = \sqrt{\frac{1}{k_b} \bar{F}}.$$

( 6)

The comparison between the evolution of the mean velocity $v$ as a function of the square root of $\bar{F}$ measured experimentally and the algebraic solution (equation ( 6)) is presented in Figure 5. The active drag coefficient $k_b$ is the fit parameter.

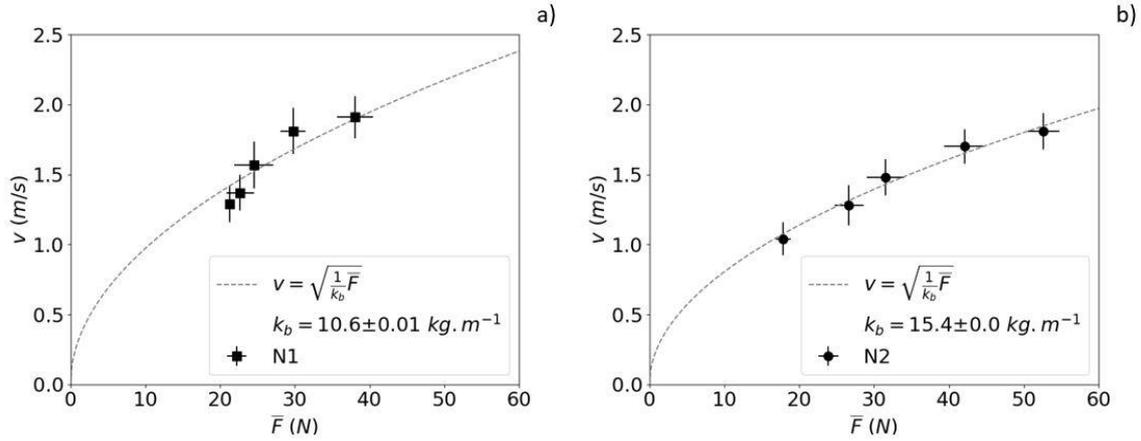

*Figure 5 : v as a function of $\bar{F}$. a) N1, national male swimmer. b) N2, international male swimmer.*

In order to highlight the gliding regime and the maximum force regime, the data for all the 25 m are presented in a graph who express the evolution of the thrust scaled by the maximum measured thrust $\frac{\widetilde{F_p}}{F^*}$ as a function of the relative propulsion time scaled by the maximum relative propulsion time $\frac{i_{prop}}{i_{prop,max}}$. A grey scale gives an indication of the average velocity scaled by the maximum measured velocity $\frac{V}{V_{max}}$ during the 25 m. The Figure 6 presents the obtained data from the test with the modelling of the two regimes for N1 and N2. For N1 (Figure 6, a), the relative thrust $\frac{\widetilde{F_p}}{F^*}$ seems to be constant and maximal during the whole test while the relative propulsion time increases. For N2 (Figure 6, b), one can observe a first phase with a constant relative propulsion time and an increase in the relative thrust. Then a second phase appear with a constant and maximal relative thrust and an increase of the relative propulsion time.

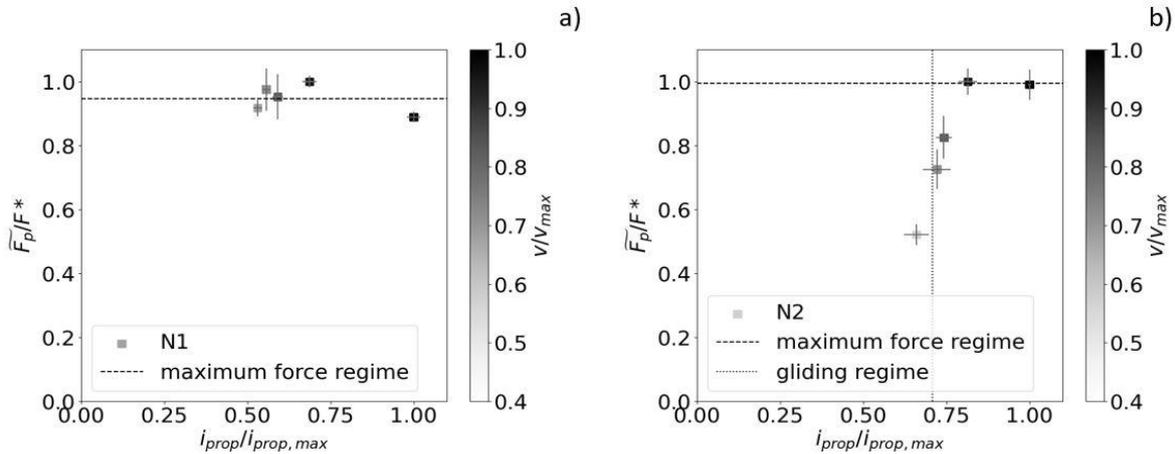

*Figure 6 : $\frac{\widetilde{F_p}}{F^*}$ as a function of $\frac{i_{prop}}{i_{prop,max}}$ with a colour scale who indicates the relative velocity $\frac{V}{V_{max}}$. a) N1, national male swimmer. b) N2, international male swimmer.*

**DISCUSSION**

The obtained results are in line with the proposed model by Carmigniani *et al.* (2020). Indeed, the graphs presented in Figure 5 show a great agreement between the algebraic solution and the experimental points for both N1 and N2. An important point to highlight is the obtained value of the active drag coefficient $k_b$ ($10.6 \pm 0.01 \ kg/m$ for N1 and $15.4 \pm 0.00 \ kg/m$ for N2) by fitting the data to the model. One can remark that these values are low compared to previous measurements of the active drag coefficient (24.3 $kg/m$ for van der Vaart *et al.* (van der Vaart, et al., 1987), 29.331 $\pm$ 2.529 $kg/m$ for Kolmogorov (Kolmogorov, 2023)). This came from the assumption done that while swimming, the propulsion is only due to arm's action and that the propulsion force has only been measured at hand's level. In other words, the experimental set up does not allow to measure the total propulsion force which lead to an under evaluation of the active drag coefficient. A way to correct that is to estimate the total propulsion force with the help of a tethered or semi-tethered test and is an ongoing work.

For the highlighting of the gliding regime and the maximum force regime (Figure 6), the two swimmers show different behaviours. N1's results suggest that he was in the maximum force regime since the first 25 m. An explanation for that can be found by looking about the smaller range of velocity explored during the test by N1 (1.29 m/s to 1.91 m/s) compared to N2 (1.04 m/s to 1.81) suggesting that the first 25 m of N1 was too fast to observe his gliding regime. N2's results present the two regimes with the gliding regime on the three first 25 m and the maximum force regime over the last two.

In conclusion, the present work is a first step towards validating this propulsion model. The algebraic solution and the data are in good agreement and the two regimes have been highlight. Future work will be to test more swimmers in order to comfort these results.

This work was funded by the Agence Nationale de la Recherche (ANR) grant number ANR-2019-STHP-0004 NePTUNE and by the EDF Fundation.